\documentclass[showpacs,preprintnumbers,amsmath,amssymb,nofootinbib]{revtex4}
\usepackage{graphicx,bm} % Include figure files
\usepackage{dcolumn}    % Align table columns on decimal point

\begin{document}

\draft
\title{Cancellation of energy-divergences and renormalizability 
in Coulomb gauge QCD within the Lagrangian formalism} 

\author{A. Ni\'{e}gawa, M. Inui and H. Kohyama} 
\address{Graduate School of Science, Osaka City University, 
Sumiyoshi-ku, Osaka 558-8585, JAPAN}
\date{Received today}

\begin{abstract}
In Coulomb gauge QCD in the Lagrangian formalism, energy divergences 
arise in individual diagrams. We give a proof on cancellation of 
these divergences to all orders of perturbation theory without 
obstructing the algebraic renormalizability of the theory. 
\hspace*{1ex} 
\end{abstract}
\maketitle 
%%%%%%%%%%%%%%%%%%%%%%%%%%%%%%%%%%%%%%%%%%%%%%
%%%% SECTION I %%%%%%%%%%%%%%%%%%%%%%%%%%%%%%%
%%%%%%%%%%%%%%%%%%%%%%%%%%%%%%%%%%%%%%%%%%%%%%
\narrowtext 
\setcounter{equation}{0}
\setcounter{section}{0}
\def\theequation{\mbox{\arabic{section}.\arabic{equation}}} 
\section{Introduction} 
In nonabelian gauge theories, the Coulomb gauge is one of the most 
important ones. The theories with this gauge are described in terms 
of physical fields, so that the unitarity is manifest. In the 
Lagrangian, second-order, formalism of Coulomb gauge perturbation 
theory, there appear, at one-loop order and above, divergent energy 
integrals of the form, 
%%%%%%%%%%%%%%%%%%%%%%%%%%%%%%%%%%%%%%%%%%%%%%%%%%%%%%%%%%%
\begin{equation}
\int \frac{d p_0}{2 \pi} \, F ({\bf p}, ... ) \, . 
\label{Einf} 
\end{equation}
%%%%%%%%%%%%%%%%%%%%%%%%%%%%%%%
Here $p_0$ is the temporal component of the four-momentum $P^\mu = 
(p_0, {\bf p})$ and \lq\lq ...'' indicates a set of external three 
momenta of the amplitude under consideration. [We use capital 
letter $Q^\mu$ to denote a four vector $Q^\mu = (q^0, {\bf q})$ 
and $q^j$ for denoting a three vector.] 

The following results are established by now on the energy 
divergence problem. 
\begin{description}
\item{1)} In the calculation using the Hamiltonian, phase-space, 
first order, form of Feynman rules, energy-divergences like 
(\ref{Einf}) do not appear \cite{doust} {\em in the first place}. 
\item{2)} On the basis of a correspondence formula which equates 
amplitudes in a covariant gauge to those in a gauge without ghosts, 
Cheng and Tsai \cite{CT} showed that, when all relevant 
contributions are added, cancellation should occur between the 
divergences like (\ref{Einf}) in the Coulomb gauge. 
\item{3)} With the aid of an interpolating gauge, which interpolates 
between a covariant gauge and the Coulomb gauge, it was shown 
\cite{BZ} in the phase-space formalism that, when all participating 
contributions are added, cancellation occurs between the divergences 
like (\ref{Einf}). (See, also, \cite{doust}.) 
\end{description}

Furthermore, following two types of ill-defined integrals appear: 
%%%%%%%%%%%%%%%%%%%%%%%%%%%%%%%%%%%%%%%%%%%%%%%%%%%%%%%%%%%
\begin{eqnarray}
&& \int \frac{d p_0}{2 \pi} \, \frac{p_0}{p_0^2 - p^2 + i 0^+} \, 
G ({\bf p}, ... ) \, , 
\label{ill1} \\ 
&& \int \frac{d p_0}{2 \pi} \frac{d q_0}{2 \pi} \, 
\frac{p_0}{p_0^2 - p^2 + i 0^+} \, 
\frac{q_0}{q_0^2 - q^2 + i 0^+} \, H ({\bf p}, {\bf q}, ... ) \, . 
\nonumber \\ 
\label{ill2}
\end{eqnarray}
%%%%%%%%%%%%%%%%%%%%%%%%%%%%%%%
The type (\ref{ill1}) appears at one-loop order and above, and the 
type (\ref{ill2}) does at two-loop order and above. 

In Coulomb-gauge QCD, one encounters a problem of operator ordering 
in the Hamiltonian. This problem was resolved by Christ and Lee 
\cite{CL}, along the line of the earlier work of Schwinger 
\cite{schw}: The quantum Hamiltonian is different from the classical 
Hamiltonian by special terms, labeled $V_1 + V_2$. Since then, 
it has been shown that the ill-defined integrals of the form 
(\ref{ill2}) are connected \cite{doust,CT1} with these $V_1 + V_2$ 
terms, and integrals of the form (\ref{ill1}) can be set equal to 
zero \cite{CT1}. 

As mentioned above, when all relevant energy-diverging contributions 
are added, energy divergences cancel out, {\em provided} that the 
remaining three-momenta integrations are convergent. It has been 
pointed out \cite{DT} that, in dealing with the renormalization 
parts, difficulties arise in simultaneously handling the diverging 
energy integrations and renormalizing the ultraviolet (UV) 
divergences. A formal proof of cancellation of energy-divergences 
and algebraic renormalizability is given in \cite{BZ} with the aid 
of an interpolating gauge in the phase-space formalism. Recently, 
this issue is studied \cite{AT} in an example in which quark-loop 
subgraphs are inserted into the second-order gluon self-energy 
graphs. As mentioned in 1) above, in the phase-space formalism, two 
integrals over the internal energies converge, provided that the 
internal three spatial momenta are held fixed. It is found in 
\cite{AT} that, when one first computes each subgraphs and performs 
renormalization, energy-divergences re-appear in the final energy 
integrals. Thanks to the Ward identity, these energy-divergent 
contributions are cancelled out when all relevant contributions are 
added. 

In this paper, generalizing the analysis in \cite{AT} to all orders 
of perturbation theory, we give a proof of cancellation of 
energy-divergences without spoiling the renormalizability of the 
Coulomb-gauge QCD by using the Feynman rules derived from the 
Lagrangian, second-order, formalism. In dealing with the 
energy-divergences of the form (\ref{Einf}), extra $V_1 + V_2$ terms 
do not participate \cite{CT1}, and then we can use the Coulomb-gauge 
Feynman rules derived from the Lagrangian formalism. 

In Sec. II, we derive a power-counting formula for divergent energy 
integrals. In Sec. III, we show that the energy divergences, Eq. 
(\ref{Einf}), are cancelled out to all orders of perturbation 
theory. In Sec. IV, for completeness, with the help of the 
power-counting formula obtained in Sec. II, we identify the diagrams 
that yields ill-defined energy integrals (\ref{ill1}) and 
(\ref{ill2}). In Sec. V, we show that the renormalization does not 
spoil the cancellation of energy divergences. 
%%%%%%%%%%%%%%%%%%%%%%%%%%%%%%%
%%% SEC %%%%%%%%%%%%%%%%%%%%%%%
%%%%%%%%%%%%%%%%%%%%%%%%%%%%%%%
\setcounter{equation}{0}
\setcounter{section}{1}
\def\theequation{\mbox{\arabic{section}.\arabic{equation}}} 
\section{Preliminary} 
QCD in the Coulomb gauge is defined, with standard notation, by the 
effective Lagrangian density 
%%%%%%%%%%%%%%%%%%%%%%%%%%
\begin{eqnarray}
{\cal L}_{\mbox{\scriptsize{eff}}} &=& - \frac{1}{4} F^{\mu \nu}_a 
F_{\mu \nu}^a - \frac{1}{2 \alpha} (\partial_i A_a^i) (\partial_j 
A_a^j) \nonumber \\ 
&& + \partial_i \bar{\eta}_a \left( \delta_{a b} \partial^i - g 
f_{a c b} A^i_c \right) \eta_b \nonumber \\ 
&& + \bar{\psi} \left( i 
\partial\kern-0.045em\raise0.3ex\llap{/}\kern0.25em\relax - m - g 
t_a {A\kern-0.1em\raise0.3ex\llap{/}\kern0.35em\relax}_a 
\right) \psi \, , 
\label{lag}
\end{eqnarray}
%%%%%%%%%%%%%%%%%%%%%%%%%%%%%
where $t_a = \lambda_a / 2$ and $F_a^{\mu \nu} = \partial^\mu 
A_a^\nu - \partial^\nu A_a^\mu - g f_{a b c} A^\mu_b A^\nu_c$. Here 
$A_a^\mu$ denotes the gluon field and $\eta_a$ ($\bar{\eta}_a$) 
denotes the (anti)FP-ghost field. We have introduced one quark 
flavor ($\psi, \bar{\psi}$). Generalization to the case with several 
quark flavors is straightforward. Generalization to other nonabelian 
gauge theories is also straightforward. 

Throughout in the sequel, we restrict ourselves to the strict 
Coulomb gauge ($\alpha \to 0$). The propagators in the Lagrangian 
formalism may be extracted from the bilinear terms (in fields) of 
${\cal L}_{\mbox{\scriptsize{eff}}}$ in Eq. (\ref{lag}); 
%%%%%%%%%%%%%%%%%%%%%%%%%%%%%
\begin{eqnarray}
\langle \mbox{T} A_a^i (x) A_b^j (y) \rangle & 
\stackrel{F.T.}{\longrightarrow} & i \delta_{ab} 
\frac{\delta^{i j}_\perp ({\bf p})}{P^2 + i 0^+} \equiv i 
\delta_{ab} D^{i j} (P) \, , \nonumber \\ 
\langle \mbox{T} A_a^0 (x) A_b^0 (y) \rangle & 
\stackrel{F.T.}{\longrightarrow} & i \delta_{ab} \frac{1}{p^2} 
\equiv i \delta_{ab} D^{00} (p) \, , \nonumber \\ 
\langle \mbox{T} \eta_a (x) \bar{\eta}_b (y) \rangle & 
\stackrel{F.T.}{\longrightarrow} & - i \delta_{ab} \frac{1}{p^2} 
\equiv i \delta_{ab} \tilde{D} (p) \, . 
\label{denpa}
\end{eqnarray}
%%%%%%%%%%%%%%%%%%%%%%%%%%%%%%%
%%%%%%%%%%%%%%%%%%%%%%%%%%%%%%%%%%%%%%%%%%%%%%%%%%%%%%%%%%%%%%%%
%%%%%%%%%%%%%%%%%%%%%%%%%%%%%
Here, $P^2 = p_0^2 - p^2$ $(p = |{\bf p}|)$, 
$\delta_\perp^{i j} 
({\bf p}) \equiv \delta^{i j} - p^i p^j /p^2$, and \lq F.T.' stands 
for taking Fourier transformation. 
%%%%%%%%%%%%%%%%%%%%%%%%%%%%%%%%%%%%%%%%%%%%%%%%%%%%%%%%%%%%%
%%%%%%%% SEC %%%%%%%%%%%%%%%%%%%%%%%%%%%%%%%%%%%%%%%%%%%%%%%%%%%%%
%%%%%%%%%%%%%%%%%%%%%%%%%%%%%%%%%%%%%%%%%%%%%%%%%
\subsubsection*{Superficial degree of energy divergence} 
It is sufficient to deal with one particle irreducible diagrams. 
>From them, we take a particular diagram $G$. We introduce the 
following abbreviations; 
\begin{description} 
\item{i)} \lq $A$' for the spatial component of the gluon field, 
$A_a^i$, which we call the transverse gluon (tgluon) in the sequel, 
\item{ii)} \lq $0$' for the temporal component of the gluon field, 
$A^0_a$, which we call the \lq Coulomb', 
\item{iii)} \lq $G$' (\lq $\bar{G}$') for the (anti)FP-ghost, 
\item{iv)} \lq $q$' (\lq $\bar{q}$') for the (anti)quark. 
\end{description} 
We adopt the following notation to describe the diagram $G$: 
%%%%%%%%%%%%%%%%%%%%%%%%
\begin{eqnarray*} 
&& I_i = \mbox{ number of internal lines of } i \;\;\;\;\; (i = A, 
0, G, q) \, , \\ 
&& E_i = \mbox{ number of external lines of } i \\ 
&& \mbox{\hspace*{30ex}} (i = A, 0, G, q \; \mbox{and} \; \bar{q}) 
\, , \\ 
&& V_{3A}, V_{4A}, V_{A A 0}, V_{A 0 0}, V_{A A 0 0}, 
V_{A \bar{G} G}, V_{A \bar{q} q}, V_{0 \bar{q} q} \\ 
&& \mbox{\hspace*{2.5ex}} = \mbox{number of vertices indicated by 
the suffices} \, . 
\end{eqnarray*} 
%%%%%%%%%%%%%%%%%%%%%%%%%%%%
Let us find the superficial degree of energy divergence, 
$\omega (G)$, of $G$. When $\omega (G) \leq - 1$, energy integral 
converges. 

To $\omega (G)$, each loop contributes $+ 1$, each internal tgluon 
line contributes $-2$, each internal quark line contributes $-1$, 
and each \lq $AA0$' vertex contributes $+ 1$. Then, it is clearly 
%%%%%%%%%%%%%%%%%%%%%%%%%%
\begin{equation}
\omega (G) = L - 2 I_A - I_q + V_{AA0} \, . 
\label{degree}
\end{equation}
%%%%%%%%%%%%%%%%%%%%%%%%%
>From the topological structure of $G$, we have 
%%%%%%%%%%%%%%%%%%%%%%%%%%
\begin{eqnarray}
L &=& I_A + I_0 + I_G + I_q - \left( \sum_i V_i - 1 \right) 
\, , \nonumber \\ 
2 I_A + E_A &=& 3 V_{3 A} + 4 V_{4A} + 2 V_{A A0} + V_{A 00} 
\nonumber \\ 
&& + 2 V_{AA00} + V_{A \bar{G} G} + V_{A \bar{q} q} \, , \nonumber 
\\ 
2 I_0 + E_0 &=& V_{A A0} + 2 V_{A 00} + 2 V_{AA00} + V_{0 \bar{q} q} 
\, , \nonumber \\ 
2 I_q + E_{q \; \mbox{\scriptsize{and}} \; \bar{q}} &=& 
2 V_{A \bar{q} q} + 2 V_{0 \bar{q} q} \, , \nonumber \\ 
2 I_G + E_G &=& 2 V_{A \bar{G} G} \, . 
\label{iden}
\end{eqnarray}
%%%%%%%%%%%%%%%%%%%%%%%%%
In the first equation, summation is taken over all types of vertices 
in $G$, where we have used the fact that there is energy 
conservation at each vertex but there is also one overall energy 
conservation. Using the relations (\ref{iden}) in Eq. 
(\ref{degree}), we obtain 
%%%%%%%%%%%%%%%%%%%%%%%%%%
\begin{eqnarray}
\omega (G) &=& 1 + \frac{1}{2} \left( E_A - V_{A A 0} - V_{A 00} - 
2 V_{AA00} - V_{A \bar{G} G} \right) \nonumber \\ 
&& - \frac{1}{2} \left( E_G + E_0 \right) 
 \nonumber \\ 
&& 
- \frac{1}{2} \left( 5 V_{3A} + 6 V_{4A} + 3 V_{A \bar{q} q} + 
V_{0 \bar{q} q} \right) \, . 
\label{super1}
\end{eqnarray}
%%%%%%%%%%%%%%%%%%%%%%%%%
The directions of the momenta of the external lines are taken toward 
outside of the diagram $G$ (c.f., Fig. 1). Eq. (\ref{super1}) tells 
us that $\omega (G) \leq 1$. 
%%%%%%%%%%%%%%%%%%%%%%%%%%%%%%%%%%%%%%%%%%%%%%
%%% SEC %%%%%%%%%%%%%%%%%%%%%%%%%%%%%%%%%%%%%%%%%%%
%%%%%%%%%%%%%%%%%%%%%%%%%%%%%%%%%%%%%%%%%%%%%%
\section{Cancellation of $\omega (G) = 1$ energy divergences}
\setcounter{equation}{0}
\setcounter{section}{3}
\def\theequation{\mbox{\arabic{section}.\arabic{equation}}} 
%%%%%%%%%%%%%%%%%%%%%%%%%%%%%%%%%%%%%%%%%%%%%%
%%% SUBSEC %%%%%%%%%%%%%%%%%%%%%%%%%%%%%%%%%%%%%%%%%%%
%%%%%%%%%%%%%%%%%%%%%%%%%%%%%%%%%%%%%%%%%%%%%%
\subsection{Participating diagrams} 
The diagram $G$ with $\omega (G) = 1$ yields the divergent integral 
of the form (\ref{Einf}). From Eq. (\ref{super1}) we learn that the 
$\omega (G) = 1$ energy-divergence arises only when the following 
three conditions are simultaneously met: 
%%%%%%%%%%%%%%%%%%%%%%%%%%%%%%%%%%%%%%%%%%%%%%
\begin{description}
\item{(C1)} $E_A = V_{AA0} + V_{A00} + 2 V_{A A00} + V_{A \bar{G} 
G}$, 
\item{(C2)} $E_0 = E_G = 0$, 
\item{(C3)} $V_{3A} = V_{4A} = V_{A \bar{q} q} = V_{0 \bar{q} q} = 
0$. 
\end{description}
%%%%%%%%%%%%%%%%%%%%%%%%%%%%%%%%%%%%%%%
These conditions leads to the following propositions: 
%%%%%%%%%%%%%%%%%%%%%%%%%%%%%%%%%%%%%%%%
\begin{description}
\item{(P1)} Energy divergence arises only from the tgluon 
amplitudes [(C2) and (C3)]. 
\item{(P2)} In an energy-divergent tgluon amplitude, the vertices 
in (C1) above, $V_{AA0}$, $V_{A00}$, $V_{A A00}$, and 
$V_{A \bar{G} G}$, are the external vertices, i.e., 
external-tgluon lines go out from them. In particular, when one 
tgluon goes out from a vertex $V_{A A00}$, if any, in the diagram 
$G$, no energy divergence arises [(C1)]. 
\item{(P3)} Energy divergent diagrams do not have internal vertex 
[(P2) and (C3)]. 
\end{description} 
Then, 
\begin{description}
\item{(P4)} Then, energy divergence can arise only from tgluon 
one-loop diagrams, Fig. 1. 
\end{description} 

\begin{figure}[h]
\begin{center}
\includegraphics[width=4cm,clip]{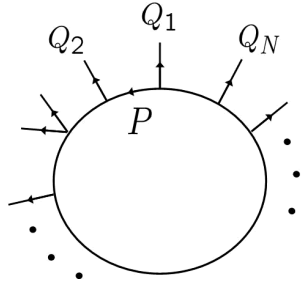}
\end{center}
\caption{One-loop N-tgluon diagram.}
\end{figure}

%%%%%%%%%%%%%%%%%%%%%%%%%%%%%%%%%%%%%%%
%%%%%%% SUBSEC %%%%%%%%%%%%%%%%%%%%%%%%%%%%%%%%
%%%%%%%%%%%%%%%%%%%%%%%%%%%%%%%%%%%%%%%
\subsection{Structure of the building blocks of the tgluon one-loop 
diagrams} 
Let us compute the amplitude for the diagram in Fig. 2(a) that is a 
part of Fig. 1: 
%%%%%%%%%%%%%%%%%%%%%%%%%%
\begin{eqnarray}
{\cal A}_{2a} &=& \left[ - g f_{bce} (p^0 + q_2^0) \right] \frac{i 
\delta^{ij}_\perp ({\bf p})}{P^2 + i 0^+} \left[ g f_{dae} (q^0_1 - 
p^0) \right] \nonumber \\ 
&=& i g^2 f_{bce} f_{dae} \left[ 1 + \frac{p^2 - (q_1^0 - q_2^0) 
p^0 - q_1^0 q_2^0}{P^2 + i 0^+} \right] \delta^{i j}_\perp ({\bf p}) 
\, . \nonumber \\ 
\label{2anano} 
\end{eqnarray}
%%%%%%%%%%%%%%%%%%%%%%%%%
The second term in the square brackets on the last line does not 
yield the $\omega (G) = 1$ energy divergence. Here and throughout in 
the following we are concerned only with the portions that 
participate in the $\omega (G) = 1$ energy divergence. Then, we have 
%%%%%%%%%%%%%%%%%%%%%%%%%%
\begin{eqnarray}
{\cal A}_{2a} & \simeq & i g^2 f_{bce} f_{dae} \left( \delta^{i j} - 
\frac{p^i p^j}{p^2} \right) \, , 
\label{fig1}
\end{eqnarray}
%%%%%%%%%%%%%%%%%%%%%%%%%
where \lq $\simeq$' indicates that the right-hand side is the 
portion of the left-hand side that leads to the energy divergent 
contribution to the one-loop amplitudes under consideration. 

\begin{figure}[h]
\begin{center}
\includegraphics[width=8cm,clip]{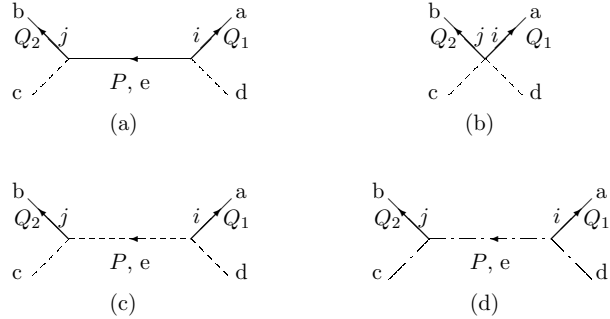}
\end{center}
\caption{Lowest order four-point diagrams, each of which is a part of Fig.1. 
Here Solid lines represent tgluons; dashed lines represent Coulombs; and dot-dashed line represents FP-ghost.}
\end{figure}

The amplitude for the diagram in Fig. 2(b) reads 
%%%%%%%%%%%%%%%%%%%%%%%%%%
\begin{eqnarray}
{\cal A}_{2b} &=& - i g^2 \left( f_{bce} f_{dae} - f_{bde} f_{ace} 
\right) \delta^{i j} \nonumber \\ 
& \equiv & {\cal A}_{2b}^{(1)} + {\cal A}_{2b}^{(2)} \, . 
\label{fig2}
\end{eqnarray}
%%%%%%%%%%%%%%%%%%%%%%%%%
We see from Eqs. (\ref{fig1}) and (\ref{fig2}) that the partial 
cancellation occurs between ${\cal A}_{2a}$ and 
${\cal A}_{2b}^{(1)}$. Similar partial cancellation occurs between 
the contribution from the diagram that is obtained from Fig. 2(a) by 
$(a, i, Q_1) \leftrightarrow (b, j, Q_2)$ and ${\cal A}_{2b}^{(2)}$. 
Thus, we have 
%%%%%%%%%%%%%%%%%%%%%%%%%%
\begin{equation}
{\cal A}_{2a} + {\cal A}_{2b}^{(1)} \simeq - i g^2 f_{bce} f_{dae} 
\frac{p^i p^j}{p^2} \;\;\; \left( \equiv {\cal A}_2 \right)\, . 
\label{fig1and2}
\end{equation}
%%%%%%%%%%%%%%%%%%%%%%%%%
With understanding that Figs. 2(a) and 2(b) are always combined into 
the form (\ref{fig1and2}), we will forget Fig. 2(b) or $V_{AA00}$ 
hereafter. 

The contribution from Fig. 2(c) reads 
%%%%%%%%%%%%%%%%%%%%%%%%%%
\begin{eqnarray}
{\cal A}_{2c} & = & \left[ g f_{bce} (2 p^j - q_2^j) \right] 
\frac{i}{p^2} \left[ g f_{dae} (2 p^i + q^i_1) \right] \nonumber \\ 
&=& 4 i g^2 f_{bce} f_{dae} \frac{p^i p^j}{p^2} \, . 
\label{fig3}
\end{eqnarray}
%%%%%%%%%%%%%%%%%%%%%%%%%
Here we have used the fact that, in the strict Coulomb gauge adopted 
here, $q_1^i \epsilon^i_r ({\bf q}_1) = 0$ $(r = 1, 2)$ and 
$q_1^i D^{i k} (Q_1) = 0$, where $\epsilon^i_r ({\bf q}_1)$ and 
$D^{i k} (Q_1)$ are, in respective order, the tgluon polarization 
vector and the propagator, Eq. (\ref{denpa}), which are to be 
attached to ${\cal A}_{2c}$. Thus, $q_1^i$ may be dropped. 
Similarly, $q_2^j$ may be dropped. 

In a similar manner, we have, for the contribution from Fig. 2(d), 
%%%%%%%%%%%%%%%%%%%%%%%%%%
\begin{eqnarray}
{\cal A}_{2d} & = & \left[ - g f_{bce} (p^j - q^j_2 ) \right] 
\frac{- i}{p^2} \left[ - g f_{dae} p^i \right] \nonumber \\ 
&=& - i g^2 f_{bce} f_{dae} \frac{p^i p^j}{p^2} \, . 
\label{fig4}
\end{eqnarray}
%%%%%%%%%%%%%%%%%%%%%%%%%
We observe that, besides the difference between the overall 
factors, the functional forms of ${\cal A}_2$, ${\cal A}_{2c}$, 
and ${\cal A}_{2d}$ are the same. We note that the integrand of 
a potentially energy divergent tgluon one-loop amplitude is 
independent of the temporal component of the loop momentum. 

Finally in this subsection, it should be emphasized that the 
one-loop diagram that includes two or more adjacent tgluon 
propagators does not yield energy divergence (c.f., Eq. 
(\ref{2anano})). 
%%%%%%%%%%%%%%%%%%%%%%%%%%%%%%%%%%%%%%%%%%%%%%%%%%%%%%%%%%%%%%
%%% SUBSUB %%%%%%%%%%%%%%%%%%%%%%%%%%%%%%%%%%%%%%%%%%%%%%%%%%%
%%%%%%%%%%%%%%%%%%%%%%%%%%%%%%%%%%%%%%%%%%%%%%%%%%
\subsection{Absence of overlapping energy divergences} 
Here, we show that no overlapping energy divergences arise. Assume 
that, in the two-loop diagram depicted in Fig. 3, both the 
left-side one-loop ($\Xi_L$) and the right-side one-loop ($\Xi_R$) 
yield energy divergence. Then, from (P4) above, four lines with 
momenta $P$, $Q$, $P'$, and $Q'$ are the tgluon lines. Furthermore, 
from the observation at the end of Sec. IIIB, the line with momentum 
$P - Q$ is the $A_0$ or Coulomb line. Then, the vertex factor for 
the vertex at which three lines with momenta $P$, $Q$, and $P - Q$ 
meet is proportional to $p_0 + q_0$. As can be seen from Eq. 
(\ref{2anano}), the \lq $p_0$' (\lq $q_0$') part participates in the 
energy divergence for $\Xi_L$ ($\Xi_R$) but {\em does not} 
participate in the energy divergence for $\Xi_R$ ($\Xi_L$). Thus, 
the energy divergence does not arise simultaneously from $\Xi_L$ and 
$\Xi_R$. 

\begin{figure}[h]
\begin{center}
\includegraphics[width=5cm,clip]{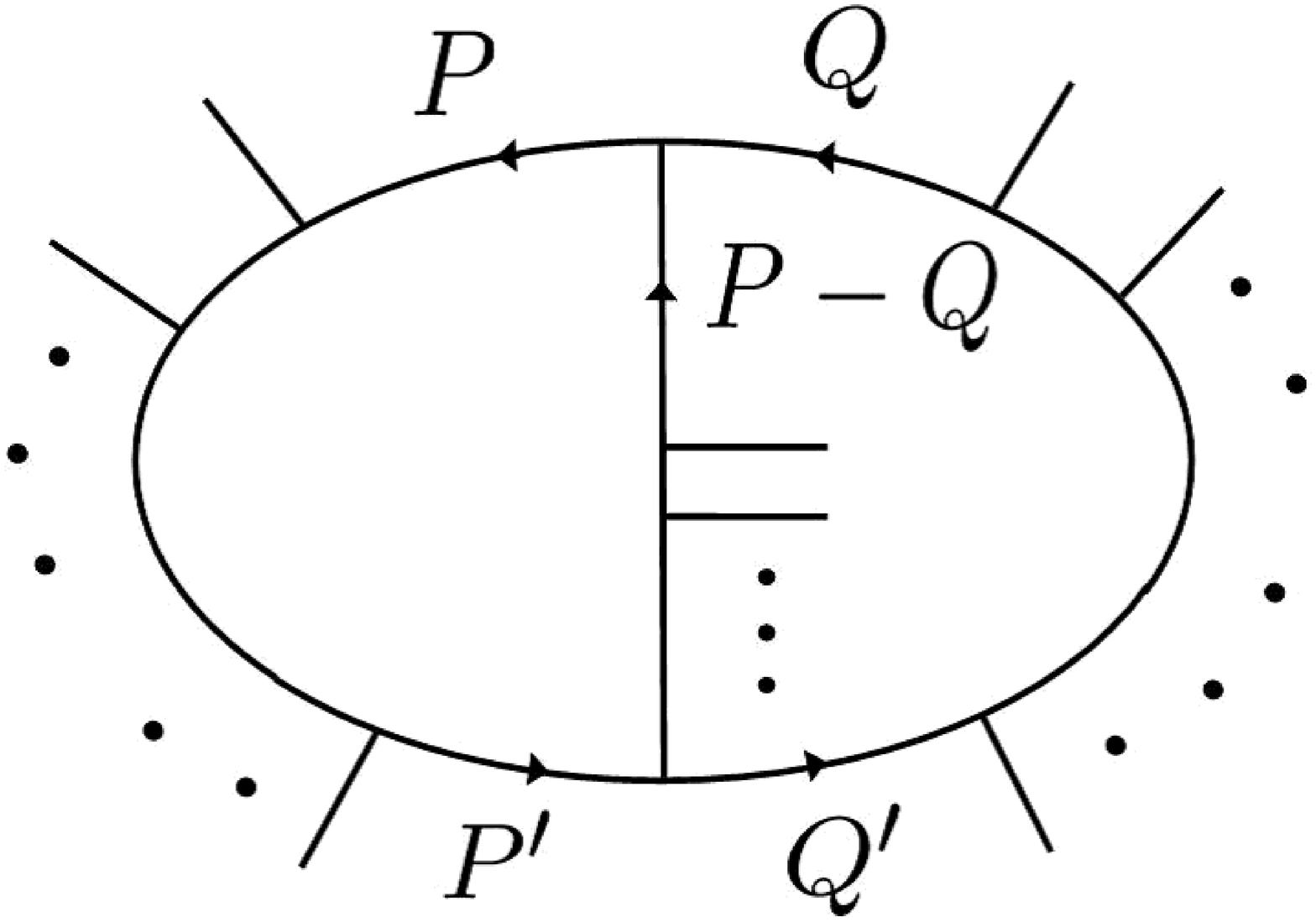}
\end{center}
\caption{Two-loop tgluon diagram.}
\end{figure}

%%%%%%%%%%%%%%%%%%%%%%%%%
%%% SUBSEC %%%%%%%%%%%%%%%%%%%%%%
%%%%%%%%%%%%%%%%%%%%%%%%%
\subsection{Cancellation of $\omega (G) = 1$ energy divergences} 
One can read off from Eqs. (\ref{2anano}) - (\ref{fig3}) the 
relative factors for the Coulomb- and the tgluon-propagators; $1 : 
- 1$, and the relative vertex factors for a $A00$- and 
$AA0$-vertices; $1 : - 1 / 2$. To represent these relative factors, 
we introduce matrices, 
%%%%%%%%%%%%%%%%%%%%%%%%%%%%%%%%%%%%%%%
\begin{equation}
\hat{\cal P} = \left( 
\begin{array}{cc}
- 1 \; & 0 \\ 
0 \; & 1 
\end{array} 
\right) \, , \;\;\;\;\;\; 
\hat{\cal V} = \left( 
\begin{array}{cc}
0 \; & - 1 / 2 \\ 
- 1 / 2 \; & 1 
\end{array} 
\right) \, . 
\label{matrix}
\end{equation}
%%%%%%%%%%%%%%%%%%%%%%%%%%%%%%%%%%%%%%%
Here the first rows and columns correspond to \lq $A$' and the 
second rows and columns to \lq $0$'. $\hat{\cal V}_{1 1} = 0$ comes 
from the fact remarked at the end of Sec. IIIB. Let 
${\cal A}_N^{(G)}$ be the sum of $N$-point tgluon amplitudes for 
Fig. 1, where the loop consists of Coulomb and/or tgluons lines. 
Then we obtain for the $\omega (G) = 1$ energy-divergent 
contribution to ${\cal A}_N^{(G)}$, 
%%%%%%%%%%%%%%%%%%%%%%%%%%%%%%%%%%%%%%%
%%%%%%%%%%%%%%%%%%%%%%%%%%%%%%%%%%%%%%%
%%%%%%%%%%%%%%%%%%%%%%%%%%%%%%%%%%%%%%%
\begin{equation}
{\cal A}_N^{(G)} \simeq \left[ \mbox{Tr} \left( \hat{\cal P} 
\hat{\cal V} \right)^N \right] {\cal A}_N^{(G 0)} \, , 
\label{oogon}
\end{equation}
%%%%%%%%%%%%%%%%%%%%%%%%%%%%%%%%%%%%%%%
where ${\cal A}_N^{(G 0)}$ is the contribution from Fig. 1, where 
all $N$ propagators are the Coulomb ones. Note that, for $N = 2$, a 
symmetry factor $1 / 2$ is necessary, which is included in 
${\cal A}_N^{(G)}$ and ${\cal A}_N^{(G 0)}$. Through mathematical 
induction, we obtain 
%%%%%%%%%%%%%%%%%%%%%%%%%%%%%%%%%%%%%%%%%%%%%%%%%%%%%%%%%
\begin{equation}
\left( \hat{\cal P} \hat{\cal V} \right)^N = 2^{- N} \left( 
\begin{array}{cc}
- N + 1 \; & N \\ 
- N \; & N + 1 
\end{array}
\right) \, , 
\label{4.15d}
\end{equation}
%%%%%%%%%%%%%%%%%%%%%%%%%%%%%%%%%
so that 
%%%%%%%%%%%%%%%%%%%%%%%%%%%%%%%%%%%%%%%
\begin{equation}
{\cal A}_N^{(G)} \simeq 2^{1 - N} {\cal A}_N^{(G 0)} \, . 
\label{gold}
\end{equation}
%%%%%%%%%%%%%%%%%%%%%%%%%%%%%%%%%%%%%%%

The tgluon amplitude ${\cal A}_N^{(FP)}$ for the 
FP-ghost one-loop diagrams is obtained from ${\cal A}_N^{(G 0)}$ 
through the following operations (cf. Eqs. (\ref{fig3}) and 
(\ref{fig4})); a) change the sign of each propagator, b) multiply a 
factor $-1 / 2$ for each vertex, c) multiply a factor $2$, for $3 
\leq N$, that corresponds to two diagrams with opposite circulation 
of fermion number, and d) multiply $- 1$ that comes from one fermion 
loop. Thus, we obtain 
%%%%%%%%%%%%%%%%%%%%%%%%%%%%%%%%%%%%%%%
\begin{eqnarray*}
{\cal A}_N^{(FP)} & = & (-) \cdot 2 \cdot (-)^N \cdot \left( - 
\frac{1}{2} \right)^N \cdot {\cal A}_N^{(G0)} 
% \nonumber 
\\ 
&& = - 2^{1 - N} {\cal A}_N^{(G 0)} \, . 
\end{eqnarray*}
%%%%%%%%%%%%%%%%%%%%%%%%%%%%%%%%%%%%%%%
For $3 \leq N$, the factor \lq 2' in the mid-term comes from c) 
above. For $N = 2$, as mentioned after Eq. (\ref{oogon}), 
${\cal A}_2^{(G0)}$ includes a symmetry factor \lq 1/2', while, 
${\cal A}_2^{(FP)}$ does not include a factor \lq $1 / 2$'. Then, 
the factor \lq $2$' is necessary in the mid-term. Thus, we see that 
${\cal A}_N^{(G)}$ cancels ${\cal A}_N^{(FP)}$: 
%%%%%%%%%%%%%%%%%%%%%%%%%%%%%%%%%%%%%%%
\begin{equation}
{\cal A}_N \equiv {\cal A}_N^{(G)} + {\cal A}_N^{(FP)} \simeq 0 
\, . 
\label{ohp} 
\end{equation}
%%%%%%%%%%%%%%%%%%%%%%%%%%%%%%%%%%%%%%%
%%%%%%%%% SUB %%%%%%%%%%%%%%%%%%%%%%%%%
%%%%%%%%%%%%%%%%%%%%%%%%%%%%%%%%%%%%%%%
\subsection{Regularization} 
As mentioned above, ${\cal A}_N^{(G 0)}$ is independent of the 
temporal component of the loop momentum and is of the form 
(\ref{Einf}), 
%%%%%%%%%%%%%%%%%%%%%%%%%%%%%%%%%%%%%%%%%%%%%%%%%%%%%%
\begin{equation}
{\cal A}_N^{(G 0)} = \int d^3 p \int d p_0 \, F ({\bf p}, ... ) \, . 
\label{byou}
\end{equation}
%%%%%%%%%%%%%%%%%%%%%%%%%%%%%%%%%%%%%%%%%%%%%%%%%%%
Integration over $p_0$ diverges. In order to rigorously handle this 
integral we should introduce some regularization. It is desirable 
to adopt the regularization that preserves the BRST symmetry of the 
theory, so that the regularizing quantum effective action $\Gamma$ 
satisfies the Zinn-Justin equation, whose explicit form is not 
necessary for our purpose. To our best knowledge, two candidates 
are available, i.e., the interpolating gauges \cite{BZ}, and the 
split dimensional regularization \cite{lei}. 

Throughout in this paper, we adopt the interpolating gauge 
\cite{BZ}, which is defined by introducing the gauge condition, 
%%%%%%%%%%%%%%%%%%%%%%%%%%%%%%%%%%%%%%%%%%%%%%
\[ 
\xi \partial_0 A_a^0 + \partial_i A_a^i = 0 
\] 
%%%%%%%%%%%%%%%%%%%%%%%%%%%%%%%%%%%%%%%%%%%%%%%%
($\xi$ is a real parameter). The Coulomb gauge is obtained by taking 
the limit $\xi \to 0$. $\xi$ here plays a role of regularization 
of the energy-divergent integrals in the Coulomb gauge. For 
arbitrary $\xi$, $\Gamma$ satisfies the Zinn-Justin equation, which, 
in the limit $\xi \to 0$, turns out to be its Coulomb-gauge 
counterpart. 

Gluon- and ghost-propagators read 
%%%%%%%%%%%%%%%%%%%%%%%%%%%%%%%%%%%%%%%
\begin{eqnarray}
\Delta^{ij} (P) &=& \frac{1}{P^2 + i 0^+} \left[ \delta^{ij} - 
\frac{p^2 - \xi (2 - \xi) p_0^2}{(P \cdot P_\xi + i 0^+)^2} p^i p^j 
\right] \, , \nonumber \\ 
\Delta^{00} (P) &=& \frac{p^2}{(P \cdot P_\xi + i 0^+ )^2} \, , 
\nonumber \\ 
\Delta^{0i} (P) &=& \Delta^{i0} = \frac{\xi p_0 p^i}{(P \cdot 
P_\xi + i 0^+ )^2} \, , \nonumber \\ 
\tilde{\Delta} (P) &=& \frac{1}{P \cdot P_\xi + i 0^+} = 
\frac{P \cdot P_\xi}{p^2} \Delta^{00} (P) \, , 
\label{obaQ} 
\end{eqnarray}
%%%%%%%%%%%%%%%%%%%%%%%%%%%%%%%%%%%%%%%
where $P_\xi \equiv (\xi p_0, {\bf p})$. 

In the following, we only keep the terms that turn out to the 
$\omega = 1$ energy-divergent ones, and we ignore the terms that 
tend to $0$ in the limit $\xi \to 0$. In place of Eqs. 
(\ref{fig1and2}), (\ref{fig3}), and (\ref{fig4}) we have, in 
respective order, 
%%%%%%%%%%%%%%%%%%%%%%%%%%%%%%%%%%%%%%%
\begin{eqnarray}
{\cal A}_2 & \simeq & - i g^2 f_{bce} f_{dae} \frac{p^2 - \xi (2 - 
\xi) p_0^2}{(P \cdot P_\xi + i 0^+)^2} p^i p^j \, , \\ 
{\cal A}_{2c} & \simeq & 4 i g^2 f_{bce} f_{dae} \frac{p^2}{(P 
\cdot P_\xi + i 0^+)^2} p^i p^j \, , 
\label{fig33}
\\ 
{\cal A}_{2d} & \simeq & i g^2 f_{bce} f_{dae} \frac{\xi p_0^2 - 
p^2}{(P \cdot P_\xi + i 0^+)^2} p^i p^j \, . 
\label{fig44}
\end{eqnarray}
%%%%%%%%%%%%%%%%%%%%%%%%%%%%%%%%%%%%%%%
For obtaining ${\cal A}_{2c}$ for Fig. 2(c), we have used the fact 
that $q_1^i \epsilon^i_r ({\bf q}_1) = O (\xi)$ $(r = 1, 2)$ and 
$q_1^i D^{i k} (Q_1) = O (\xi)$, where $\epsilon^i_r ({\bf q}_1)$ 
and $D^{i k} (Q_1)$ are, in respective order, the tgluon 
polarization vector and the propagator, which are to be attached to 
${\cal A}_{2c}$. Then, $q_1^i$ that was present in Eq. (\ref{fig33}) 
(cf. Eq. (\ref{fig3})) turn out to be of $O (\xi)$ and lead to 
vanishing contribution in the limit $\xi \to 0$. Similarly, $q_2^j$ 
that was present in Eqs. (\ref{fig33}) and (\ref{fig44}) can be 
ignored. 

\begin{figure}[h]
\begin{center}
\includegraphics[width=9cm,clip]{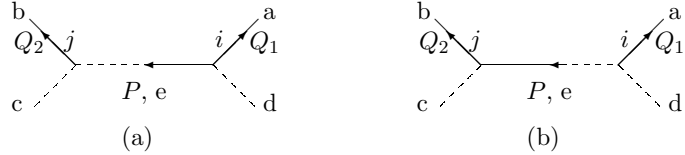}
\end{center}
\caption{Lowest order four-point diagrams, each of which is a part of Fig.1. }
\end{figure}

In the interpolating gauge, two diagrams as depicted in Fig. 4 also 
participate. Corresponding amplitudes read 
%%%%%%%%%%%%%%%%%%%%%%%%%%%%%%%%%%%%%%%
\begin{eqnarray*}
{\cal A}_{4a} & = & \left[ g f_{bce} (2 p - q_2)^j \right] 
\frac{i \xi p_0 p^i}{(P \cdot P_\xi + i 0^+)^2} \left[ g f_{dae} 
(q_1^0 - p^0) \right] 
% \nonumber 
\\ 
& \simeq & - 2 i g^2 f_{bce} f_{dae} \frac{\xi p_0^2}{(P \cdot P_\xi 
+ i 0^+ )^2} p^i p^j 
% \nonumber 
\\ 
& \simeq & {\cal A}_{4b} \, . 
% \label{fig88}
\end{eqnarray*}
%%%%%%%%%%%%%%%%%%%%%%%%%%%%%%%%%%%%%%%
Diagrams as depicted in Fig. 5 do not yield energy divergence in 
the limit $\xi \to 0$. 

\begin{figure}[h]
\begin{center}
\includegraphics[width=9cm,clip]{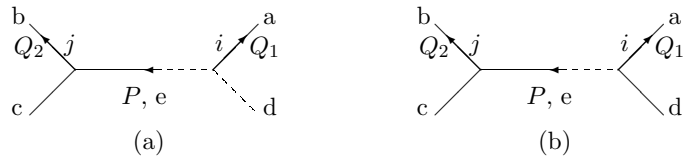}
\end{center}
\caption{Lowest order four-point diagrams. }
\end{figure}

Let us make a change of variable, $p_0 \to \tilde{p}_0 / 
\sqrt{\xi}$. Then, ignoring the terms that lead to 
vanishing contribution in the limit $\xi \to 0$, we have 
%%%%%%%%%%%%%%%%%%%%%%%%%%%%%%%%%%%%%%%
\begin{eqnarray}
{\cal A}_2 & \simeq & - i g^2 f_{bce} f_{dae} (p^2 - 2 
\tilde{p_0}^2) {\cal D}^{ij} (\tilde{P}) \, , \\ 
{\cal A}_{2c} & \simeq & 4 i g^2 f_{bce} f_{dae} p^2 {\cal D}^{ij} 
(\tilde{P}) \, , \\ 
{\cal A}_{2d} & \simeq & i g^2 f_{bce} f_{dae} (\tilde{p}_0^2 - 
p^2) {\cal D}^{ij} (\tilde{P}) \, , 
\label{oba} \\ 
{\cal A}_{4a} & \simeq & {\cal A}_{4b} \simeq - 2 i g^2 f_{bce} 
f_{dac} \tilde{p}_0^2 {\cal D}^{ij} 
(\tilde{P}) \, , 
\end{eqnarray}
%%%%%%%%%%%%%%%%%%%%%%%%%%%%%%%%%%%%%%%
where $\tilde{P} \equiv (\tilde{p}_0, {\bf p})$ and 
%%%%%%%%%%%%%%%%%%%%%%%%%%%%%%%%%%%%%%%
\[
{\cal D}^{ij} (\tilde{P}) \equiv \frac{p^i p^j}{(\tilde{P}^2 + i 
0^+)^2} \, . 
\]
%%%%%%%%%%%%%%%%%%%%%%%%%%%%%%%%%%%%%%%

>From these equations, we obtain, in place of Eq. (\ref{matrix}), 
for $\hat{\cal P}$ and $\hat{\cal V}$, 
%%%%%%%%%%%%%%%%%%%%%%%%%%%%%%%%%%%%%%%
\begin{equation}
\hat{\cal P}_i = \left( 
\begin{array}{cc}
2 \tilde{p}_{i0}^2 / p^2_i - 1 \; & \tilde{p}_{i0}^2 / p^2_i \\ 
\tilde{p}_{i0}^2 / p^2_i \; & 1 
\end{array} 
\right) \, , \; 
\hat{\cal V} = \left( 
\begin{array}{cc}
0 \;\; & - 1 / 2 \\ 
- 1 / 2 \; & 1 
\end{array} 
\right) \, . 
\label{matrix1}
\end{equation}
%%%%%%%%%%%%%%%%%%%%%%%%%%%%%%%%%%%%%%%
The contributions that turn out to $\omega (G) = 1$ 
energy-divergent contributions in the limit $\xi \to 0$ is given by 
(cf. Eq. (\ref{oogon})), 
%%%%%%%%%%%%%%%%%%%%%%%%%%%%%%%%%%%%%%%
\begin{equation}
{\cal A}_N^{(G)} \simeq \frac{1}{\sqrt{\xi}} \int \frac{d 
\tilde{p}_0}{2 \pi} \mbox{Tr} \left[ \prod_{l = 1}^N \left( 
\hat{\cal P}_l \hat{\cal V} \right) \right] A_N^{(G 0)} + O 
(\xi^{1 / 2}) \, , 
\label{oogon1}
\end{equation}
%%%%%%%%%%%%%%%%%%%%%%%%%%%%%%%%%%%%%%%
where ${\cal A}_N^{(G0)} = (2 \pi \sqrt{\xi})^{- 1} \int d 
\tilde{p}_0 A_N^{(G0)}$. Through mathematical induction, one can 
show that 
%%%%%%%%%%%%%%%%%%%%%%%%%%%%%%%%%%%%%%%%%%%%%%%%%%%%%%%%%
\[ 
\mbox{Tr} \left[ \prod_{l = 1}^N \left( \hat{\cal P}_l \hat{\cal V} 
\right)^N \right] = 2^{1 - N} \prod_{l = 1}^N \frac{p_l^2 - 
\tilde{p}_{l0}^2}{p_l^2} \, . 
\]
%%%%%%%%%%%%%%%%%%%%%%%%%%%%%%%%%
Then, Eq. (\ref{oogon1}) turns out to 
%%%%%%%%%%%%%%%%%%%%%%%%%%%%%%%%%%%%%%%
\begin{equation}
{\cal A}_N^{(G)} \simeq \frac{2^{1 - N}}{\sqrt{\xi}} \int 
\frac{d \tilde{p}_0}{2 \pi} \left[ \prod_{l = 1}^N \frac{p_l^2 - 
\tilde{p}_{l0}^2}{p_l^2} \right] A_N^{(G 0)} + O (\xi^{1 / 2}) 
\, . 
\label{gold1}
\end{equation}
%%%%%%%%%%%%%%%%%%%%%%%%%%%%%%%%%%%%%%%

The amplitude for the FP-ghost one-loop diagram is obtained 
similarly as in Sec. IIID using Eq. (\ref{obaQ}) or Eq. (\ref{oba}): 
%%%%%%%%%%%%%%%%%%%%%%%%%%%%%%%%%%%%%%%
\begin{equation}
{\cal A}_N^{(FP)} \simeq - \frac{2^{1 - N}}{\sqrt{\xi}} \int 
\frac{d \tilde{p}_0}{2 \pi} \left[ \prod_{l = 1}^N \frac{p_l^2 - 
\tilde{p}_{l0}^2}{p_l^2}\right] A_N^{(G 0)} + O (\xi^{1 / 2}) 
\, . 
\label{gold2}
\end{equation}
%%%%%%%%%%%%%%%%%%%%%%%%%%%%%%%%%%%%%%%
Then, we see that $O (\xi^{- 1 / 2})$ contributions to 
${\cal A}_N^{(G)}$ and ${\cal A}_N^{(FP)}$ are cancelled out, and, 
in the limit $\xi \to 0$, we have 
%%%%%%%%%%%%%%%%%%%%%%%%%%%%%%%%%%%%%%%
\[
%\begin{equation}
{\cal A}_N \equiv {\cal A}_N^{(G)} + {\cal A}_N^{(FP)} \simeq 0 
\, . 
% \label{ohp1} 
%\end{equation}
\]
%%%%%%%%%%%%%%%%%%%%%%%%%%%%%%%%%%%%%%%
Thus, Eq. (\ref{ohp}) gets a sound foundation. 

It should be emphasized that the diagrams in Fig. 4, which are 
absent in the strict Coulomb gauge, participate here. 

Analysis with the split dimensional regularization leads to the same 
result, which we do not reproduce. 
%%%%%%%%%%%%%%%%%%%%%%%%%%%%%%%%%%%%%%%%%%%%%%
%%% SEC %%%%%%%%%%%%%%%%%%%%%%%%%%%%%%%%%%%%%%%%%%%
%%%%%%%%%%%%%%%%%%%%%%%%%%%%%%%%%%%%%%%%%%%%%%
\setcounter{equation}{0}
\setcounter{section}{3}
\def\theequation{\mbox{\arabic{section}.\arabic{equation}}} 
\section{Ill-defined integrals with $\omega (G) = 0$}
In this section, for completeness, we briefly mention the diagrams 
with $\omega (G) = 0$. From Eq. (\ref{super1}) together with the 
observation in the last section, we see that $\omega (G) = 0$ energy 
divergences and ill-defined integrals arise from the following 
diagrams (see, also, \cite{doust}): 
%%%%%%%%%%%%%%%%%%%%%%%%%%
\begin{description}
\item{1)} Gluon one-loop amplitudes, of which a number of external 
Coulomb field is at most one. 
\item{2)} $q$-$\bar{q}$-tgluon one-loop amplitudes of the type 
as shown in Fig. 6. 
\item{3)} Tgluon two-loop amplitudes (Fig. 3). 
\item{4)} Two-loop amplitudes of the type as depicted in Fig. 7. 
\end{description}

\begin{figure}[btp]
  \begin{center}
  \begin{minipage}{.45\linewidth}
  \includegraphics[width=6cm,clip]{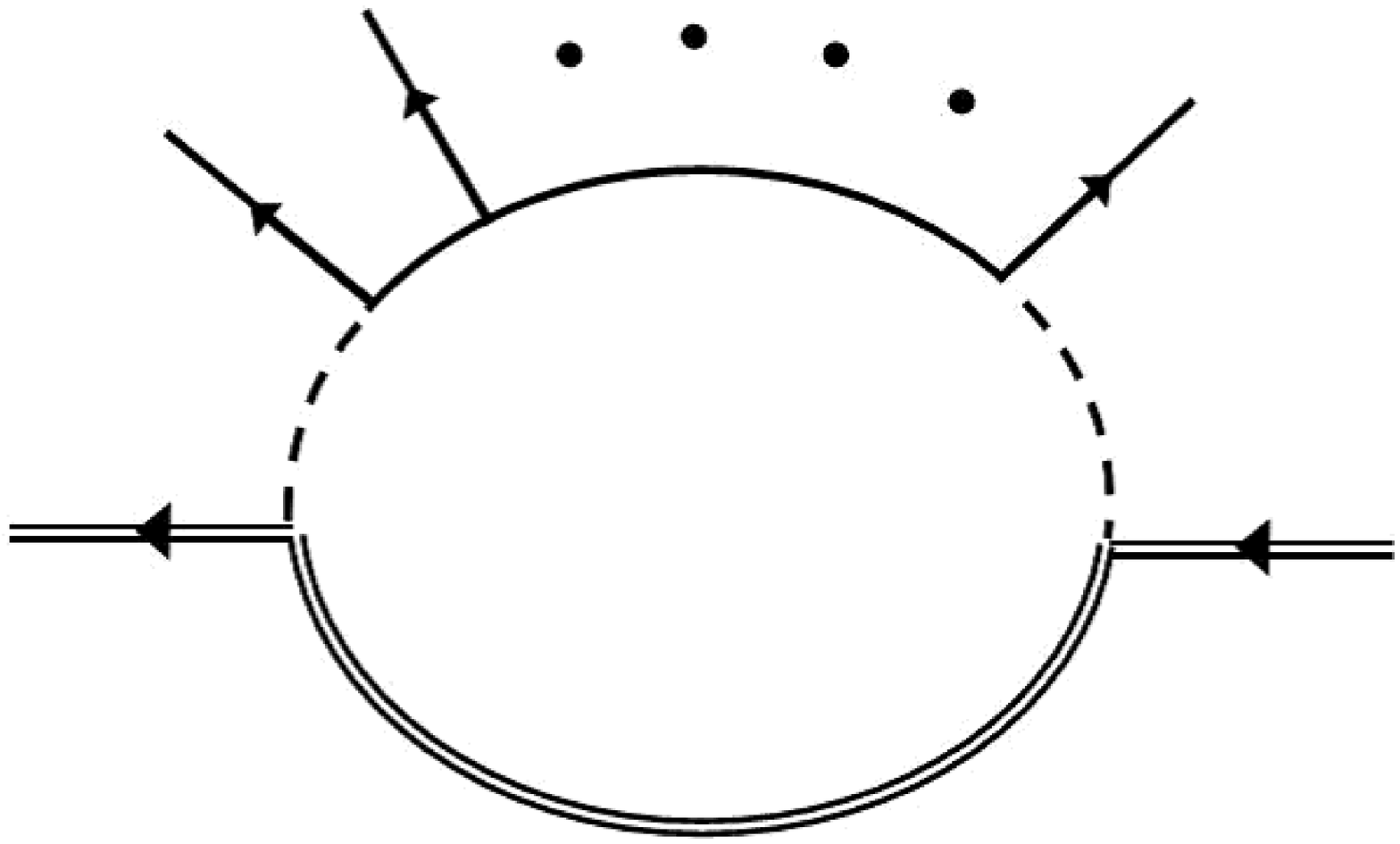}
  \caption{One-loop $q$-$\bar{q}$-tgluon(s) diagram. The double line represents a quark. 
Here each solid line that constitutes the loop is a Coulomb or a tgluon.}
  \end{minipage}
  \hspace{2.0pc}
  \begin{minipage}{.45\linewidth}
  \includegraphics[width=6cm,clip]{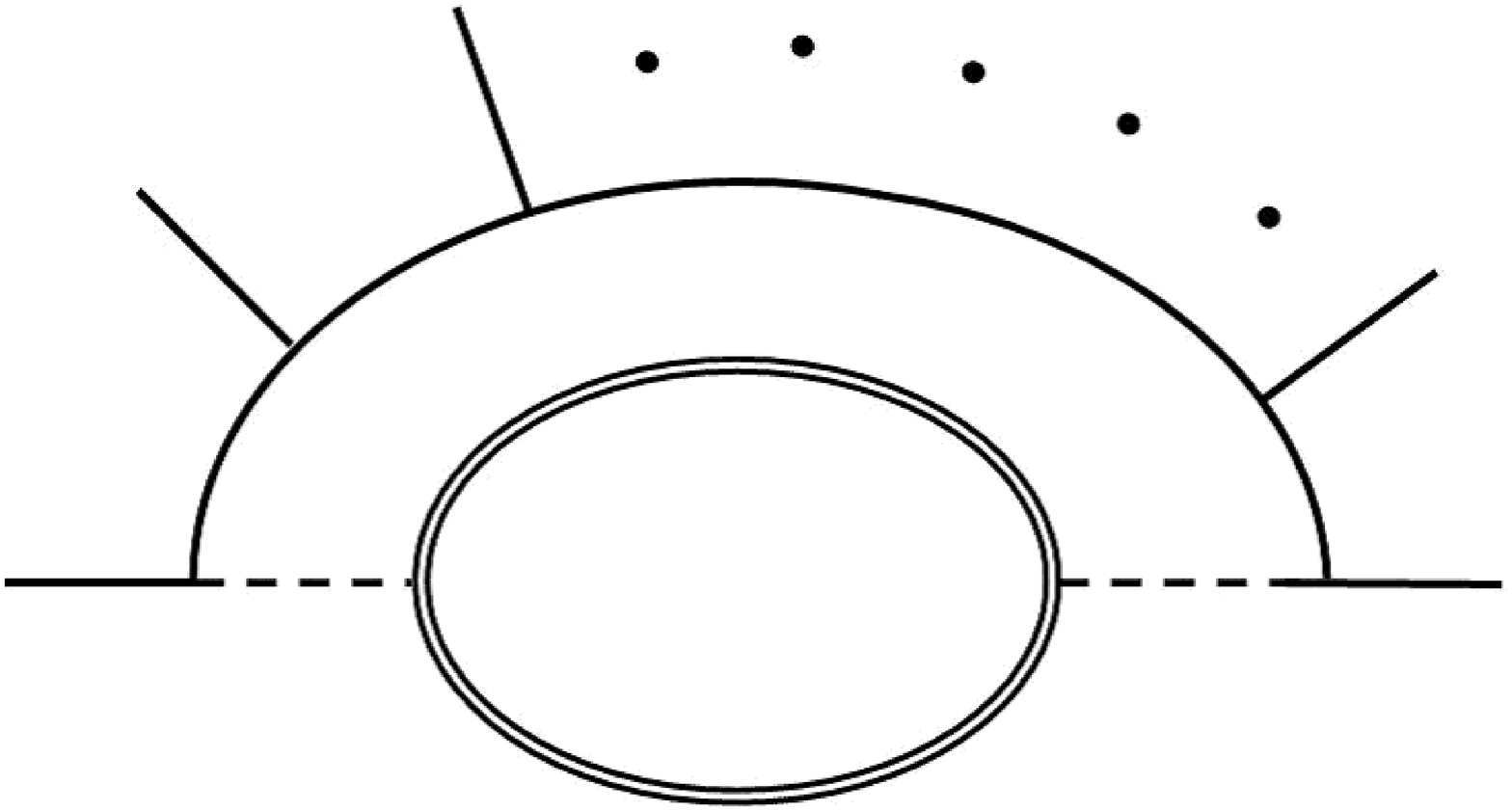}
  \caption{Two-loop tgluon diagram. The double line represents a quark. 
each solid line that constitutes the loop is a Coulomb or a tgluon.} 
  \end{minipage}
  \end{center}
  \end{figure}

Now let us inspect Fig. 3. Cutting the line with momentum $P$ (or 
$Q$ or $P - Q$), we obtain an one-loop diagram. If $\omega = 1$ 
energy divergence arises from this one-loop diagram, as has been 
proved in the previous section, it is cancelled out together with 
its relatives. Then, from Fig. 3, integrals of the forms 
(\ref{ill1}) and (\ref{ill2}) emerge, which are ill-defined ones. 
The integrals of the form (\ref{ill1}) emerge from 1) - 3), and 
the integrals of the form (\ref{ill2}) emerge from 3) and 4). As 
mentioned in Sec. I, the form (\ref{ill1}) can be set equal to zero, 
and the integrals of the form (\ref{ill2}) are connected with $V_1 + 
V_2$ terms of Christ and Lee \cite{CL}. 

This is, however, not the end of the story. As a matter of fact, 
newcomers enter into the stage through renormalization, i.e., the 
renormalization counterterms. In the next section, we deal with this 
issue. 
%%%%%%%%%%%%%%%%%%%%%%%%%%%%%%%%%%%%%%%
%%% SEC %%%%%%%%%%%%%%%%%%%%%%%%%%%%%%%%%%%%
%%%%%%%%%%%%%%%%%%%%%%%%%%%%%%%%%%%%%%%
\setcounter{equation}{0}
\setcounter{section}{4}
\def\theequation{\mbox{\arabic{section}.\arabic{equation}}} 
\section{Compatibility of cancellation of energy divergences and 
renormalizability}
We adopt the minimal subtraction scheme in dimensional 
regularization and introduce the counter Lagrangian density 
$\delta {\cal L}_{\mbox{\scriptsize{eff}}}$ that includes different 
renormalization constants. The effective QCD Lagrangian density 
${\cal L}_{\mbox{\scriptsize{eff}}}$ with addition of $\delta 
{\cal L}_{\mbox{\scriptsize{eff}}}$ in the strict Coulomb 
gauge reads \cite{nie} 
\widetext 
%%%%%%%%%%%%%%%%%%%%%%%%%%%%%%%%%%%%%%%%
\begin{eqnarray} 
{\cal L}_{\mbox{\scriptsize{eff}}} + \delta 
{\cal L}_{\mbox{\scriptsize{eff}}} &=& - \frac{Z_{31}}{2} 
\left( \partial_\mu A_{aj} \partial^\mu A_a^j 
- \partial_i A_a^j \partial_j A_a^i \right) + Z_{32} \partial_0 
A_a^i \partial_i A_a^0 - \frac{Z_{33}}{2} \partial_i A_a^0 
\partial^i A_a^0 \nonumber \\ 
&& - \frac{1}{2 \alpha} (\partial_i A_a^i) (\partial_j A_a^j) 
\nonumber \\ 
&& + g Z_{11} f_{abc} (\partial_i A_j^a) A_b^i A_c^j 
+ g Z_{31} f_{abc} \partial^0 A_a^i A_b^0 A_{c i} + g Z_{32} f_{abc} 
\partial^i A^0_a A_{bi} A_c^0 \nonumber \\ 
&& - \frac{g^2}{4} Z_{41} f_{abc} f_{ade} A_{bi} A_{cj} A^i_d A^j_e 
- \frac{g^2}{2} Z_{31} f_{abc} f_{ade} A^0_b A_{ci} A^0_d A^i_e 
\nonumber \\ 
&& + \tilde{Z}_3 (\partial_i \bar{\eta}_a ) \partial^i \eta_a 
- g f_{abc} (\partial_i \bar{\eta}_a) A_b^i \eta_c \nonumber \\ 
&& + Z_2 \bar{\psi} \left( i 
\partial\kern-0.045em\raise0.3ex\llap{/}\kern0.25em\relax - (m - 
\delta m) \right) \psi - g Z_1 \bar{\psi} t_a A_a^j \gamma_j 
\psi - g Z_2 \bar{\psi} t_a A_a^0 \gamma^0 \psi \, , 
\label{dog} 
\end{eqnarray} 
%%%%%%%%%%%%%%%%%%%%%%%%%%%%%%%%%%%%%
\narrowtext 

\noindent 
where the limit $\alpha \to 0$ is understood to be taken. $Z$'s and 
$\tilde{Z}$'s in Eq. (\ref{dog}) obey the Slavnov-Taylor identities 
in the narrow sense, 
%%%%%%%%%%%%%%%%%%%%%%%%%%%%%%%%%%%%%
\begin{equation}
\frac{Z_{31}}{Z_{11}} = \tilde{Z}_3 = \frac{Z_{11}}{Z_{41}} = 
\frac{Z_{32}}{Z_{31}} = \frac{Z_{33}}{Z_{32}} = \frac{Z_2}{Z_1} \, . 
\label{STyo}
\end{equation} 
%%%%%%%%%%%%%%%%%%%%%%%%%%%%%%%%%%%%%%%%
The counter Lagrangian $\delta {\cal L}_{\mbox{\scriptsize{eff}}}$ 
introduces new vertices, which we call counter vertices. Repeating 
the argument in Secs. II and III on the superficial degree of energy 
divergence, we find that the energy divergence arises from the 
diagrams that are obtained from each energy-diverging diagram $G_i$ 
$(i = 1, 2, ...)$ treated so far by inserting the counter vertices 
in all possible ways. Let ${\cal A}_i^{(j)}$ $(j = 1, 2, ...)$ be 
the amplitude for thus obtained $j$th diagram, which may be computed 
from the Lagrangian density ${\cal L}_{\mbox{\scriptsize{eff}}} + 
\delta {\cal L}_{\mbox{\scriptsize{eff}}}$ in Eq. (\ref{dog}). 
Summing over all ${\cal A}_i^{(j)}$'s, we obtain ${\cal A} = 
\sum_i \left({\cal A}_i^{(0)} + \sum_j {\cal A}_i^{(j)} \right)$, 
where ${\cal A}_i^{(0)}$ is the amplitude for $G_i$.  

We now show that ${\cal A}$ is free from energy divergence. Eq. 
(\ref{dog}) tells us that the propagators and the vertex factors 
extracted from ${\cal L}_{\mbox{\scriptsize{eff}}} + \delta 
{\cal L}_{\mbox{\scriptsize{eff}}}$ are obtained from their bare 
counterparts through the following replacements: 
%%%%%%%%%%%%%%%%%%%%%%%%%%
\begin{eqnarray}
& D^{ij} (P) \rightarrow Z_{31}^{- 1} D^{ij} (P) \, , 
\;\;\;\;\;\;\;\;\;\;\; 
& D^{00} (p) \rightarrow Z_{33}^{- 1} D^{00} (p) \, , 
\nonumber \\ 
& \tilde{D} (p) \rightarrow \tilde{Z}_3^{- 1} \tilde{D} (p) \, , 
\;\;\;\;\;\;\;\;\;\;\;
& V^{i j k}_{abc} \rightarrow Z_{11} V^{ijk}_{abc} \, , \nonumber \\ 
& V^{i j 0}_{abc} \rightarrow Z_{31} V^{ij0}_{abc} \, , 
\;\;\;\;\;\;\;\;\;\;\;
& V^{i 0 0}_{abc} \rightarrow Z_{32} V^{i00}_{abc} \, , \nonumber \\ 
& V^{ijkl}_{abcd} \rightarrow Z_{41} V^{ijkl}_{abcd} \, , 
\;\;\;\;\;\;\;\;\;\;\;
& V^{ij00}_{abcd} \rightarrow Z_{31} V^{ij00}_{abcd} 
\, , \nonumber \\ 
& \tilde{V}^i_{abc} \rightarrow \tilde{V}^{i}_{abc} \, .\qquad \qquad \; \,   & {} 
\label{okikae}
\end{eqnarray}

%%%%%%%%%%%%%%%%%%%%%%%%%%%%%
where $V^{i j k}_{abc}$ is the three-tgluon vertex factor, 
$V^{i j 0}_{abc}$ is the $A_a^i A_b^j A_c^0$-vertex factor, 
$\tilde{V}^i_{abc}$ is the $\bar{\eta}_a A_b^i \eta_c$-vertex 
factor, and so on. Explicit form of the quark propagator is not 
necessary for our purpose. 

We first show that the key equations (\ref{fig1and2}) and 
(\ref{ohp}) remain to hold. From Eq. (\ref{okikae}), we see that 
${\cal A}_{2a}$, Eq. (\ref{fig1}), and ${\cal A}_{2b}^{(1)}$, Eq. 
(\ref{fig2}), turn out to be 
%%%%%%%%%%%%%%%%%%%%%%%%%%%%%%%%%
\[
%\begin{equation}
{\cal A}_{2 a} \longrightarrow \frac{Z_{31}^2}{Z_{31}} 
{\cal A}_{2 a} = Z_{31} {\cal A}_{2 a} \, , 
%\label{iro}
%\end{equation}
\]
%%%%%%%%%%%%%%%%%%%%%%%%%%%%%%%%%%%%%%%
and  
%%%%%%%%%%%%%%%%%%%%%%%%%%%%%%%%%
\[
%\begin{equation}
{\cal A}_{2 b}^{(1)} \longrightarrow Z_{31} {\cal A}_{2 b}^{(1)} 
\, , 
%\label{hani}
%\end{equation}
\]
%%%%%%%%%%%%%%%%%%%%%%%%%%%%%%%%%%%%%%%
respectively. Then, the same partial cancellation as for 
${\cal A}_{2 a}$ and ${\cal A}_{2 b}^{(1)}$ in Sec. III occurs, and 
we obtain, in place of Eq. (\ref{fig1and2}), 
%%%%%%%%%%%%%%%%%%%%%%%%%%
\begin{equation} 
{\cal A}_{2a} + {\cal A}_{2b}^{(1)} \longrightarrow 
Z_{31} \left( {\cal A}_{2a} + {\cal A}_{2b}^{(1)} \right) = 
Z_{31} {\cal A}_2 \, . 
\end{equation}
%%%%%%%%%%%%%%%%%%%%%%%%%
Taking ${\cal A}_N^{(G 0)}$ as the reference amplitude, we have, in 
place of the matrices $\hat{\cal P}$ and $\hat{\cal V}$ in Eq. 
(\ref{matrix}), 
%%%%%%%%%%%%%%%%%%%%%%%%%%%%%%%%%%%%%%%
\begin{eqnarray*}
&& \hat{\cal P} = \left( 
\begin{array}{cc}
- Z_{31}^{- 1} Z_{33} \; & 0 \\ 
0 \; & 1 
\end{array} 
\right) \, , 
%\nonumber 
\\ 
&& \hat{\cal V} = \left( 
\begin{array}{cc}
0 \; & - Z_{31} Z_{32}^{- 1} / 2 \\ 
- Z_{31} Z_{32}^{- 1} / 2 \; & 1 
\end{array} 
\right) \, . 
%\label{matrix3}
\end{eqnarray*}
%%%%%%%%%%%%%%%%%%%%%%%%%%%%%%%%%%%%%%%
Using the identities (\ref{STyo}), we have, in place of Eq. 
(\ref{4.15d}), 
%%%%%%%%%%%%%%%%%%%%%%%%%%%%%%%%%%%%%%%
\[
\left( \hat{\cal P} \hat{\cal V} \right)^N = 2^{- N} \left( 
\begin{array}{cc}
- N + 1 \; & N (Z_{32} / Z_{31}) \\ 
- N (Z_{31} / Z_{32}) \; & N + 1 
\end{array}
\right) \, . 
\]
%%%%%%%%%%%%%%%%%%%%%%%%%%%%%%%%%
Then, we see that the relation (\ref{gold}) holds as it is: 
%%%%%%%%%%%%%%%%%%%%%%%%%%%%%%%%%%%%%%%
\[
%\begin{equation}
{\cal A}_N^{(G)} \simeq 2^{1 - N} {\cal A}_N^{(G 0)} \, . 
%\label{gold5}
%\end{equation}
\]
%%%%%%%%%%%%%%%%%%%%%%%%%%%%%%%%%%%%%%%
The tgluon amplitude for the FP-ghost one-loop diagrams is obtained 
using Eq. (\ref{okikae}),  
%%%%%%%%%%%%%%%%%%%%%%%%%%%%%%%%%%%%%%%
\[
{\cal A}_N^{(FP)} = - 2^{1 - N} \frac{1}{\tilde{Z}_3^N} 
\left[ \left( \frac{Z_{33}}{Z_{32}} \right)^N {\cal A}_N^{(G 0)} 
\right] = - 2^{1 - N} {\cal A}_N^{(G 0)} \, , 
\]
%%%%%%%%%%%%%%%%%%%%%%%%%%%%%%%%%%%%%%%
where use has been made of Eq. (\ref{STyo}). Thus, 
${\cal A}_N^{(G)}$ cancels ${\cal A}_N^{(FP)}$: 
%%%%%%%%%%%%%%%%%%%%%%%%%%%%%%%%%%%%%%%
\begin{equation}
{\cal A}_N^{(G)} + {\cal A}_N^{(FP)} \simeq 0 \, . 
\label{5.9d} 
\end{equation}
%%%%%%%%%%%%%%%%%%%%%%%%%%%%%%%%%%%%%%%
%%%%%% SUBSUB %%%%%%%%%%%%%%%%%%%%%%%%%%%%%%%%%
%%%%%%%%%%%%%%%%%%%%%%%%%%%%%%%%%%%%%%%
\subsubsection*{Proof of Eq. (\ref{5.9d}) with the aid of the 
regularization by an interpolating gauge} 
The effective QCD Lagrangian density with addition of the counter 
Lagrangian density in the interpolating gauge, $\xi \partial_0 A_a^0 
+ \partial_i A_a^i = 0$, reads 
\widetext 
%%%%%%%%%%%%%%%%%%%%%%%%%%%%%%%%%%%%%%%%
\begin{eqnarray} 
{\cal L}_{\mbox{\scriptsize{eff}}} + \delta 
{\cal L}_{\mbox{\scriptsize{eff}}} &=& - \frac{Z_{31}}{2} \left( 
\partial_\mu A_{aj} \partial^\mu A_a^j - \partial_i A_a^j \partial_j 
A_a^i \right) + Z_{32} \partial_0 A_a^i \partial_i A_a^0 - 
\frac{Z_{33}}{2} \partial_i A_a^0 \partial^i A_a^0 \nonumber \\ 
&& - \frac{1}{2 \alpha} (\tilde{\partial}_\mu A_a^\mu) 
(\tilde{\partial}_\nu A_a^\nu) \nonumber \\ 
&& + g Z_{11} f_{abc} (\partial_i A_j^a) A_b^i A_c^j + g Z_{12} 
f_{abc} \partial^0 A_a^i A_b^0 A_{c i} + g Z_{13} f_{abc} \partial^i 
A^0_a A_{bi} A_c^0 \nonumber \\ && - \frac{g^2}{4} Z_{41} f_{abc} 
f_{ade} A_{bi} A_{cj} A^i_d A^j_e - \frac{g^2}{2} Z_{42} f_{abc} 
f_{ade} A^0_b A_{ci} A^0_d A^i_e \nonumber \\ 
&& + \tilde{Z}_{31} (\partial_i \bar{\eta}_a ) \partial^i 
\eta_a + \xi \tilde{Z}_{32} (\partial_0 \bar{\eta}_a ) 
\partial^0 \eta_a - g f_{abc} (\tilde{\partial}_\mu \bar{\eta}_a) 
A_b^\mu \eta_c \nonumber \\ 
&& + Z_2 \bar{\psi} \left( i 
\partial\kern-0.045em\raise0.3ex\llap{/}\kern0.25em\relax - (m - 
\delta m) \right) \psi - g Z_{\psi 11} \bar{\psi} t_a A_a^j \gamma_j 
\psi - g Z_{\psi 12} \bar{\psi} t_a A_a^0 \gamma^0 \psi \, , 
\label{dog1} 
\end{eqnarray} 
%%%%%%%%%%%%%%%%%%%%%%%%%%%%%%%%%%%%%%%%
where $\tilde{\partial}_\mu \equiv (\xi \partial_0, \nabla)$ and 
$\alpha \to 0$ is understood to be taken. $Z$'s and $\tilde{Z}$'s 
obey 
%%%%%%%%%%%%%%%%%%%%%%%%%%%%%%%%%%%%%%%%%%%
\begin{eqnarray} 
\frac{Z_{31}}{Z_{11}} &=& \tilde{Z}_{31} = \frac{Z_{11}}{Z_{41}} = 
\frac{Z_{32}}{Z_{12}} = \frac{Z_{33}}{Z_{13}} = 
\frac{Z_2}{Z_{\psi 11}} \, \; (\equiv D) \, , 
\label{A18d} \\ 
\frac{Z_{31}}{Z_{12}} &=& \tilde{Z}_{32} = \frac{Z_{12}}{Z_{42}} = 
\frac{Z_{32}}{Z_{13}} = \frac{Z_2}{Z_{\psi 12}} \, \; (\equiv D') 
\, . 
\label{STyo1}
\end{eqnarray} 
%%%%%%%%%%%%%%%%%%%%%%%%%%%%%%%%%%%%%%%%
These equations are derived in a similar manner as in \cite{nie}. A 
few comments are in order. 
\begin{itemize}
\item Diagrammatic analysis shows that $\bar{\eta} A^\mu \eta$ 
three-point function is UV finite, as it does in the Landau 
gauge. 
\item In the strict Coulomb-gauge limit, $\xi \to 0$, $D' = 1$. 
\end{itemize}

Gluon- and ghost- propagators can be read off from Eq. (\ref{dog1}): 
%%%%%%%%%%%%%%%%%%%%%%%%%%%%%%%%%%%%%
\begin{eqnarray}
\Delta^{ij} & = & \frac{1}{Z_{31}} \frac{1}{P^2 + i 0^+} 
\left[\delta^{ij} - \frac{p^2 - 2 \xi (Z_{32} / Z_{33}) p_0^2 + 
\xi^2 (Z_{31} / Z_{33}) p_0^2}{(p^2 - (Z_{31} / Z_{32}) \xi p_0^2 + 
i 0^+)^2} \, p^i p^j \right] \, , \nonumber \\ 
&=& \frac{1}{Z_{31}} \frac{1}{P^2 + i 0^+} \left[\delta^{ij} - 
\frac{p^2 - 2 \xi (D' / D) p_0^2 + \xi^2 (D' / D)^2 p_0^2}{(p^2 - 
(D' / D) \xi p_0^2 + i 0^+)^2} \, p^i p^j \right] \, , \nonumber \\ 
\Delta^{00} &=& \frac{Z_{31}}{Z_{32}^2} \frac{p^2}{(p^2 - (D' / D) 
\xi p_0^2 + i 0^+)^2} \, , \nonumber \\ 
\Delta^{0i} &=& \Delta^{i 0} = \frac{Z_{31}}{Z_{32}^2} 
\frac{\xi p_0 p^i}{(p^2 - (D' / D) \xi p_0^2 + i 0^+)^2} \, , 
\nonumber \\ 
\tilde{\Delta} &=& \frac{1}{\xi \tilde{Z}_{32} p_0^2 - 
\tilde{Z}_{31} p^2 + i 0^+} = \frac{1}{\tilde{Z}_{31}} 
\frac{1}{\xi (\tilde{Z}_{32} / \tilde{Z}_{31}) p_0^2 - p^2 + i 0^+} 
\nonumber \\ 
&=& \frac{1}{\tilde{Z}_{31}} 
\frac{(D' / D) \xi p_0^2 - p^2}{(p^2 - (D' / D) \xi p_0^2 - i 
0^+)^2} \, , 
\label{prode} 
\end{eqnarray}
%%%%%%%%%%%%%%%%%%%%%%%%%%%%%%%%%%%%
\narrowtext 

\noindent 
where use has been made of Eqs. (\ref{A18d}) and (\ref{STyo1}). 
Through the same manner as above leading to Eq. (\ref{matrix1}), we 
obtain for $\hat{\cal P}_i$ and $\hat{\cal V}$, 
%%%%%%%%%%%%%%%%%%%%%%%%%%%%%%%%%%%%%%%%%%%%%%%%%%%%%%%%%%%
\begin{eqnarray*}
\hat{\cal P}_i &=& \left( 
\begin{array}{cc}
( D / D' )^2 
[ 2 (D' / D) \tilde{p}_{i0}^2 / p_i^2 - 1] \;\;\;\; & 
\tilde{p}_{i 0}^2 / p_i^2 \\ 
\tilde{p}_{i 0}^2 / p_i^2 \;\; & 1 
\end{array}
\right) \, , 
%\nonumber 
\\ 
\hat{\cal V} & = & 
\left( 
\begin{array}{cc}
0 \;\;\;\; & - Z_{12} / (2 Z_{13}) \\ 
- Z_{12} / (2 Z_{13}) \;\; & 1 
\end{array}
\right) 
% \nonumber 
\\ 
& = & \left( 
\begin{array}{cc}
0 \;\;\;\; & - D' / (2 D) \\ 
- D' / (2 D) \;\; & 1 
\end{array}
\right) \, , 
%\label{obaba}
\end{eqnarray*} 
%%%%%%%%%%%%%%%%%%%%%%%%%%%%%%%%%%%%%%%%%%%%%%%%%%%%
where $\tilde{p}_{i0} = \sqrt{\xi} p_{i0}$. Matrix multiplication 
yields 
%%%%%%%%%%%%%%%%%%%%%%%%%%%%%%%%%%%%%%%%%%%%%%%%%%%%
\[
\hat{\cal P}_i \hat{\cal V} = \frac{1}{2} \left( 
\begin{array}{cc} 
A_i - 1 \;\;\; & D / D' \\ 
- D' / D \;\; & A_i + 1 
\end{array}
\right) \, , 
\]
%%%%%%%%%%%%%%%%%%%%%%%%%%%%%%%%%%%%%%%%%%%%%%%%%%%%
where 
%%%%%%%%%%%%%%%%%%%%%%%%%%%%
\[
A_i = 1 - (D' / D) \tilde{p}_{i 0}^2 / p_i^2 \, . 
\]
%%%%%%%%%%%%%%%%%%%%%%%%%%%%%
Mathematical induction yields 
%%%%%%%%%%%%%%%%%%%%%%%%%%%%
\begin{eqnarray*}
\prod_{i = 1}^N \left( \hat{\cal P}_i \hat{\cal V} \right) & = & 
\frac{1}{2^N} \left( 
\begin{array}{cc}
\alpha_N^{(+)} \;\;\; & \beta_N \\ 
\gamma_N \;\; & \alpha_N^{(+)} 
\end{array}
\right) \, , \\ 
\alpha_N^{(\pm)} &=& \left[ \prod_{i = 1}^N A_i \right] \left( 1 \mp 
\sum_{i = 1}^N \frac{1}{A_i} \right) \, , \\ 
\beta_N & = & - \left( \frac{D}{D'} \right)^2 \gamma_N = 
\frac{D}{D'} \left[ \prod_{i = 1}^N A_i \right] 
\sum_{i = 1}^N \frac{1}{A_i} \, . 
\end{eqnarray*}
%%%%%%%%%%%%%%%%%%%%%%%%%%%%%
Then, in place of Eq. (\ref{gold1}), we obtain 
%%%%%%%%%%%%%%%%%%%%%%%%%%%%%%%%%%%%%%%
\[
{\cal A}_N^{(G)} \simeq \frac{2^{1 - N}}{\sqrt{\xi}} \int 
\frac{d \tilde{p}_0}{2 \pi} \left[ \prod_{l = 1}^N \frac{p_l^2 - (D' 
/ D) \tilde{p}_{l0}^2}{p_l^2} \right] A_N^{(G 0)} \, . 
\] 
%%%%%%%%%%%%%%%%%%%%%%%%%%%%%%%%%%%%%%%
Using Eqs. (\ref{dog1}) - (\ref{STyo1}) and (\ref{prode}), we 
obtain, in place of Eq. (\ref{gold2}), 
%%%%%%%%%%%%%%%%%%%%%%%%%%%%%%%%%%%%%%%
\begin{eqnarray*} 
{\cal A}_N^{(FP)} & \simeq & - \frac{2^{1 - N}}{\sqrt{\xi}} \int 
\frac{d \tilde{p}_0}{2 \pi} \frac{1}{\tilde{Z}_{31}^N} \left[ 
\prod_{l = 1}^N \frac{p_l^2 - (D' / D) \tilde{p}_{l0}^2}{p_l^2} 
\right] \\ 
&& \mbox{\hspace*{10ex}} \times \left( \frac{Z_{33}}{Z_{13}} 
\right)^N A_N^{(G 0)} \nonumber \\ 
& = & - \frac{2^{1 - N}}{\sqrt{\xi}} \int 
\frac{d \tilde{p}_0}{2 \pi} \left[ \prod_{l = 1}^N \frac{p_l^2 - (D' 
/ D) \tilde{p}_{l0}^2}{p_l^2} \right] A_N^{(G 0)} \, . 
\end{eqnarray*}
%%%%%%%%%%%%%%%%%%%%%%%%%%%%%%%%%%%%%%%
Then cancellation occurs between ${\cal A}_N^{(G)}$ and 
${\cal A}_N^{(FP)}$, and then, removing the regulator $\xi \to 0$, 
we have 
%%%%%%%%%%%%%%%%%%%%%%%%%%%%%%%%%%%%%%%
\[ 
{\cal A}_N^{(G)} + {\cal A}_N^{(FP)} \simeq 0 \, . 
\] 
%%%%%%%%%%%%%%%%%%%%%%%%%%%%%%%%%%%%%%%
Thus, Eq. (\ref{5.9d}) gets a sound foundation. 
%%%%%%%%%%%%%%%%%%%%%%%%%%%%%%%%%%%%%%%%%%%%%%%%%%%%%%%%%%%%%%%%
%%% SUBSUB %%%%%%%%%%%%%%%%%%%%%%%%%%%%%%%%%%%%%%%%%%%%%%%%
%%%%%%%%%%%%%%%%%%%%%%%%%%%%%%%%%%%%%%%%%%%%%%%%%%%
\subsubsection*{Compatibility of cancellation of energy divergences 
and renormalizability}
What we have shown above is that, when the renormalization 
counterterms are included to all orders of perturbation theory, 
energy-divergent contributions are cancelled out. Then, by 
expanding $Z$'s and $\tilde{Z}$'s in powers of $g^2$, 
energy-divergent contributions are cancelled order by order in 
perturbation theory. 

Armed with this proposition, we are now in a position to show that 
the cancellation of energy divergences are compatible with the 
renormalizability of the theory. 

We are concerned with the tgluon $N$-point one-loop amplitudes 
${\cal A}_N$ $(N = 2, 3, ...)$. Let ${\cal A}_N^{(M)}$ be the $O 
(g^{N + M})$ contribution to ${\cal A}_N$ and $G_i$ ($i = 1, 2, 
...$) be the set of one-loop diagrams that contributes to 
${\cal A}_N^{(0)}$. For regularizing the energy divergence, we 
employ the interpolating gauge, and to handle the UV divergence 
we employ the dimensional regularization by continuing the spacetime 
dimensionality from $4$ to $d$. Then, for $\xi \neq 0$ and $d \neq 
4$, all contributions to ${\cal A}_N$ are free from energy- and 
UV-divergences. 
\begin{description}
\item{(I) $O (g^N)$:} As shown in Sec. III, ${\cal A}_N^{(0)}$ 
is free from energy divergence, i.e., finite in the limit $\xi \to 
0$. For $5 \leq N$, ${\cal A}_N^{(0)}$ is finite in the limit $d \to 
4$ (UV finite). For $2 \leq N \leq 4$, ${\cal A}_N^{(0)}$ is written 
in the form, 
%%%%%%%%%%%%%%%%%%%%%%%%%%%%%%%%%%%%%%%%%%%%%%%%%%%%
\[
{\cal A}_N^{(0)} = \frac{1}{\sqrt{\xi}} {\cal F}_N^{(0)} + 
{\cal G}_N^{(0)} + O (\xi^{1 / 2}) \, . 
\]
%%%%%%%%%%%%%%%%%%%%%%%%%%%%%%%%%%%%%%%%%%%%%%%%%%%%%%%%%%%%%
Algebraic renormalizability (for arbitrary $\xi$) described above in 
conjunction with Eqs. (\ref{dog1}) - (\ref{STyo1}) indicates that 
${\cal F}_N^{(0)}$ and ${\cal G}_N^{(0)}$ are finite in the limit 
$d \to 4$ (UV finite). Furthermore, thanks to the above proof of 
cancellation of energy divergences, we have ${\cal F}_N^{(0)} = 0$, 
so that, by removing the regulator $\xi \to 0$, we see that 
${\cal A}_N^{(0)}$ $(= {\cal G}_N^{(0)})$ is free from energy 
divergence and UV finite. 
%%%%%%%%%%%%%%%%%%%%%%%%%%%%%%%%%%%%%%%%%%%%%%%%%%%%
\item{(II) $O (g^{N + 2})$:} Through inserting a single one-loop 
renormalization part into each diagram $G_i$ in all possible ways, 
we obtain a set of diagrams $G_i^{(j)}$ $(j = 1, 2, ...)$. Let 
$\left({\cal A}_N^{(2)} \right)_i^{(j)}$ be the amplitude for 
$G_i^{(j)}$: ${\cal A}_N^{(2)} = \sum_{i, \, j} 
\left({\cal A}_N^{(2)} \right)_i^{(j)}$. As mentioned in Sec. IIIC, 
no overlapping energy divergence arises in $\left({\cal A}_N^{(2)} 
\right)_i^{(j)}$. Energy divergences and/or UV divergences that 
arise from one-loop subdiagrams have already been managed at the 
first stage (I). Then, ${\cal A}_N^{(2)}$ may be written in the 
form, 
%%%%%%%%%%%%%%%%%%%%%%%%%%%%%%%%%%%%%%%%%%%%%%%%%%%%
\[
{\cal A}_N^{(2)} = \frac{1}{\sqrt{\xi}} {\cal F}_N^{(2)} + 
{\cal G}_N^{(2)} + O (\xi^{1 / 2}) \, . 
\]
%%%%%%%%%%%%%%%%%%%%%%%%%%%%%%%%%%%%%%%%%%%%%%%%%%%%%%%%%%%%%
The same argument as above in (I) leads ${\cal A}_N^{(2)} 
\stackrel{\xi \to 0}{\longrightarrow} {\cal G}_N^{(2)}$ to be UV 
finite. 
%%%%%%%%%%%%%%%%%%%%%%%%%%%%%%%%%%%%%%%%%%%%%%%%%%%%
\item{(III) Higher orders:} With the aid of the 
Bogoliubov-Parasiuk-Hepp-Zimmerman (BPHZ) prescription 
\cite{bogo}, one can proceed to higher orders and verify that 
${\cal A}_N^{(M)}$ is free from energy divergence and UV finite. 
\end{description}

Finally, it is worth mentioning that, in the course of perturbative 
computation, if infrared and/or mass singularities arise, we 
introduce small mass for gluons. It is well known that, in any 
reaction rate, the cancellation occur between them \cite{ln}. 
%%%%%%%%%%%%%%%%%%%%%%%%%%%%%%%
%%%% ACK %%%%%%%%%%%%%%%%%%%%%%%%%%%%%%%%%%
%%%%%%%%%%%%%%%%%%%%%%%%%%%%
\section*{Acknowledgments}
A. N. is supported in part by the Grant-in-Aid for Scientific 
Research [(C)(2) No. 17540271] from the Ministry of Education, 
Culture, Sports, Science and Technology, Japan, No.(C)(2)-17540271. 
%%%%%%%%%%%%%%%%%%%%%%%%%%%%%%%%%%%%%%%%%%%%%%%%%%%%%%%%%%%%
%%% Ref %%%%%%%%%%%%%%%%%%%%%%%%%%%%%%%%%%%%%%%%%%%%%%%%%%%%
%%%%%%%%%%%%%%%%%%%%%%%%%%%%%%%%

\end{document}